\documentclass[12pt]{article}
\usepackage{latexsym, amsmath, amssymb}
\newtheorem{thm}{Theorem}

\newtheorem{lem}{Lemma}
\newtheorem{prop}{Proposition}
\newtheorem{defn}{Definition}
\newtheorem{rem}{Remark}
\newcommand{\norm}[1]{\left\Vert#1\right\Vert}
\newcommand{\set}[1]{\left\{#1\right\}}

\newcommand{\To}{\longrightarrow}
\newcommand{\hilb}{\mathcal{H}}
\newcommand{\proj}{\mathcal{P}}
\newcommand{\gr}[2]{\ensuremath{Gr_#1(#2)}}
\newcommand{\image}[1]{\textrm{Im}(#1)}
\newcommand{\kernel}[1]{\textrm{ker}(#1)}
\newcommand{\trace}[1]{\textrm{trace}(#1)}

\newcommand{\scalar}[2]{\langle#1,#2\rangle}
\newcommand{\gamomd}{\Rightarrow}

\newcommand{\ssup}[2]{#1^{^#2}}
\newcommand{\can}[2]{\Pi_#1(#2)}
\newcommand{\canp}[2]{\Pi^\perp_#1(#2)}
\newenvironment{prf}{\textbf{Proof.}}{$\blacksquare$\bigskip\\}
\begin{document}
\title{Hamiltonian Systems on Complex Grassmann Manifold.
Holonomy and Schr\"{o}dinger Equation}
\author{Zakaria Giunashvili\\
\small{Department of Theoretical Physics,}\\
\small{Institute of Mathematics, Georgian Academy of Sciences}\\
\small{Tbilisi, Georgia}\\
}
\date{\today}
\maketitle
\begin{abstract}
Differential geometric structures such as the principal bundle for the canonical vector bundle
on a complex Grassmann manifold, the canonical connection form on this bundle, the canonical symplectic
form on a complex Grassmann manifold and the corresponding dynamical systems are investigated. The
Grassmann manifold is considered as an orbit of the co-adjoint action and the symplectic form is
described as the restriction of the canonical Poisson structure on a Lie coalgebra. The holonomy
of the connection on the principal bundle over Grassmannian and its relation with
\textbf{Berry phase} is considered and investigated for the integral curves of Hamiltonian
dynamical systems.
\end{abstract}
\newpage
\tableofcontents
\newpage
\section{Introduction}
The main purposes of this paper are
\begin{enumerate}
\item
to review all the \emph{canonical} differential geometric structures on Grassmann manifold of the
finite-dimensional complex subspaces of a complex Hilbert space.
\item
to investigate the canonical symplectic structure, the corresponding dynamical systems and the
holonomies of the integral curves of these dynamical systems for complex Grassmann manifolds.
\end{enumerate}
We start from the differential structure on the complex Grassmann manifold and then investigate
the connection on the canonical principal bundle and its holonomies. It turns out that the
canonical connection on this principal bundle and its holonomy plays an important role in the
theory of the geometric quantum computations (see \cite{Fujii-0}, \cite{Fujii-1}, \cite{Fujii-2},
\cite{Zan-Ras-1}, \cite{Zan-Ras-2}, \cite{Zan-Ras-3}), which indicates the universality of the
Grassmann manifold not only in the theory of fiber bundles, but also in the geometric theory of the
quantum computations. The central point in these topics is the pure geometric fact, that any
unitary transformation of the fiber of the canonical fiber bundle over a complex Grassmann manifold can
be obtained as the holonomy of a closed curve. But for physical reasons, we need not any closed
curve on a Grassmann manifold, but only ''physical`` ones; i.e., the curves that are the integral
curves for a Hamiltonian dynamical system. For an integral curve of a Hamiltonian system on a
Grassmann manifold, the corresponding curve on the total space of the canonical bundle is the
solution of the corresponding Schr\"{o}dinger equation. But the latter, in general, is not a
horisontal curve, which is the lifting the curve on the base manifold. It ''becomes`` horizontal
in \textbf{adiabatic limit}. We investigate the geometry of such curves and its relation with
Berry phase.
\section[The Differential Structure on Complex Grassmann Manifold]
{The Differential Structure and Coordinate Systems on Complex Grassmann Manifold}
Let $\hilb$ be a complex Hilbert space, infinite or finite -dimensional, as required.
For any positive integer number $m$ let us denote by \gr{m}{\hilb} the set of all
$m$-dimensional complex subspaces of the space $\hilb$. It is clear that any $m$-dimensional
subspace $X$ of the Hilbert space $\hilb$ can be uniquely defined by the corresponding operator
of orthogonal projection $\proj(X):\hilb\To X\subset\hilb$. The operator $\proj(X)$ is
characterised by the properties: $\proj(X)^*=\proj(X)$ and $\trace{\proj(X)}=\dim(X)$.

Let us denote by $\proj_m(\hilb)$ the set
\begin{equation}\label{projections}
\proj_m(\hilb)=\set{P:\hilb\To\hilb\,|\,P^2=P,\,P^*=P,\,\trace{P}=m}
\end{equation}
We have that the mapping $\proj:\gr{m}{\hilb}\To\proj_m(\hilb)$ is a bijection
(see \cite{Fujii-1}, \cite{Fujii-2}, \cite{Fujii-3}) and defines an injection of the set
\gr{m}{\hilb} in the vector space of Hermitian operators on the Hilbert space $\hilb$. This
injection induces a topology and a differential structure on the set \gr{m}{\hilb}. The set
\gr{m}{\hilb}, together with this differential structure is known as the Grassmann manifold of
$m$-dimensional complex subspaces of the Hilbert space $\hilb$. As it follows from the above
discussion, further we can identify the following two objects: $\proj_m(\hilb)$ and
$\gr{m}{\hilb}$.

Any fixed element $X\in\gr{m}{\hilb}$ defines a mapping
$$
\begin{array}{l}
\Gamma_X:Hom(X,X^\perp)\To\gr{m}{\hilb}\\
\\
\textrm{where, for }\varphi\in Hom(X,X^\perp):\;\Gamma_X(\varphi)=\set{x+\varphi(x)\,|\,x\in X}
\end{array}
$$
In other words, the subspace corresponding to the linear mapping
$X\stackrel{\varphi}{\To}X^\perp$ is the graph of the mapping $\varphi$. In some cases, we shall
ommit the subscript $_X$ in the expression $\Gamma_X$ and write just $\Gamma$. This mapping is an
injection of the vector space $Hom(X,X^\perp)$ into the set \gr{m}{\hilb} and its image is
the subset
$$
\set{Y\in\gr{m}{\hilb}\,|\,Y\cap X^\perp=\set{0}}\equiv\mathcal{W}_X
$$
It follows that varying the element $X\in\gr{m}{\hilb}$ we can cover the Grassmann manifold by
the open subsets $\mathcal{W}_X$. Actually, it is sufficient to take only a finite number of
of the elements $X\in\gr{m}{\hilb}$ to cover the total manifold by the sets of the type $\mathcal{W}_X$.

Hence, we can state that a pair $(Hom(X,X^\perp),\Gamma_X)$, for $X\in\gr{m}{\hilb}$,
defines a local coordinate system in the neighborhood of the point $X$ (see \cite{Milnor-Stasheff}).

Now, let us find the expression for the diffeomorphism
$$
\proj:\gr{m}{\hilb}\To\proj_m(\hilb)
$$
in the above local coordinate system. In other words, the problem is to find the coordinate
expression for the the following composition map
$$
\Phi_X:Hom(X,X^\perp)\stackrel{\Gamma_X}{\To}\gr{m}{\hilb}\stackrel{\proj}{\To}\proj_m(\hilb)
$$
For any $f\in Hom(X,X^\perp)$ the graph of the operator $-f^*\in Hom(X^*,X)$ is the
orthogonal subspace of the space $\Gamma_X(f)$. Indeed, for $x\in X$ and $u\in X^\perp$ we have the
following
$$
\begin{array}{l}
\scalar{x+f(x)}{u-f^*(u)}=\underbrace{\scalar{x}{u}}_0-\scalar{x}{f^*(u)}+\scalar{f(x)}{u}-
\underbrace{\scalar{f(x)}{f^*(u)}}_0=\\
=\scalar{f(x)}{u}-\scalar{x}{f^*(u)}=0\textrm{ (by the definition of dual operator).}
\end{array}
$$
For $f\in Hom(X,X^\perp)$ consider the following operator on the Hilbert space $\hilb$:
$$
\widetilde{f}=
\left[
\begin{array}{cc}
\mathbf{1} & -f^*\\
           &     \\
f          & \mathbf{1}
\end{array}
\right]:X\oplus X^\perp\To X\oplus X^\perp
$$
This operator is automorphism and it is easy to verify that $\widetilde{f}(X)=\Gamma(f)$ and
$\widetilde{f}(X^\perp)=\Gamma(-f^*)=\Gamma(f)^\perp$. It is clear that if $P_U$ is the
projection operator on a subspace $U\subset W$corresponding to a decomposition of a vector
space $W=U\oplus V$ and $A:W\To W$ is an automorphism, then the projection operator on the
subspace $A(U)$, corresponding to the decomposition $W=A(U)\oplus A(V)$, is
$P_{A(U)}=AP_UA^{-1}$. Therefore, the projection operator, corresponding to the decomposition
$\hilb=\Gamma(f)\oplus\Gamma(f)^\perp=\widetilde{f}(X)\oplus\widetilde{f}(X^\perp)$ is
$\proj(\Gamma(f))=\widetilde{f}\proj(X)\widetilde{f}^{-1}$. The explicit expression for the
operator $\widetilde{f}^{-1}$, for the decomposition $\hilb=X\oplus X^\perp$ is
$$
\left[
\begin{array}{cc}
\mathbf{1} & -f^* \\
& \\
f          & \mathbf{1}
\end{array}
\right]^{-1}=
\left[
\begin{array}{cc}
(1+f^*f)^{-1} & f^*(1+ff^*)^{-1} \\
& \\
-f(1+f^*f)^{-1} & (1+ff^*)^{-1}
\end{array}
\right]
$$
\begin{rem}
The operator $1+f^*f:X\To X$ is invertible, because
$$
\begin{array}{l}
(1+f^*f)(x)=0\;\gamomd\;(f^*f)(x)=-x\;\gamomd\;\scalar{(f^*f)(x)}{x}=-\norm{x}^2\\
\gamomd\;\norm{f(x)}^2=-\norm{x}^2\;\gamomd\;x=0
\end{array}
$$
The same is true for the operator $1+ff^*:X^\perp\To X^\perp$.
\end{rem}
Therefore, the explicit expression for the projection operator $\proj(\Gamma(f))$, corresponding
to the decomposition $\hilb=X\oplus X^\perp$ is
\begin{equation}\label{projection_in_coordinates}
\begin{array}{l}
\proj(\Gamma(f))=
\left[
\begin{array}{cc}
\mathbf{1} & -f^* \\
           &      \\
f          & \mathbf{1}
\end{array}
\right]
\left[
\begin{array}{cc}
\mathbf{1} & 0 \\
           &   \\
0          & 0
\end{array}
\right]
\left[
\begin{array}{cc}
(1+f^*f)^{-1}   & f^*(1+ff^*)^{-1} \\
                &      \\
-f(1+f^*f)^{-1} & (1+ff^*)^{-1}
\end{array}
\right]=\\
\\
=
\left[
\begin{array}{cc}
(1+f^*f)^{-1}  & f^*(1+ff^*)^{-1}  \\
               &                   \\
f(1+f^*f)^{-1} & ff^*(1+ff^*)^{-1} \\
\end{array}
\right]
\end{array}
\end{equation}
\\
\bigskip
The action of the group $U(\hilb)$ (the group of the unitary transformations of the Hilbert space $\hilb$,
on $\hilb$) induces the action of this group on the Grassmann manifold. This action is transitive
and for any point $X\in\gr{m}{\hilb}$, the corresponding stabilizer subgroup is
$U(X)\times U(X^\perp)$. Therefore, the manifold $\gr{m}{\hilb}$ can be considered as the
homogeneous space (see, for example, \cite{Fujii-1}, \cite{Fujii-2}, \cite{Fujii-3})
$$
\gr{m}{\hilb}\cong\frac{U(\hilb)}{U(X)\times U(X^\perp)}
$$
For any unitary transformation $u:\hilb\To\hilb$, the corresponding diffeomorphism of the
Grassmann manifold $gr(u):\gr{m}{\hilb}\To\gr{m}{\hilb}$ induces a transformation of the local
coordinate systems
$$
\widetilde{u}:Hom(X,X^\perp)\To Hom(Y,Y^\perp)
$$
where $Y=u(X)$ (and therefore $Y^\perp=u(X^\perp)$). This transformation is defined by the
condition: $\Gamma(\widetilde{u}(f))=u(\Gamma(f))$. We have that
$$
u(\Gamma(f))=\set{ux+uf(x)\,|\,x\in X}=\set{y+uf(u^{-1}y)\,|\,y\in u(X)\equiv Y}
$$
which implies that $u(\Gamma(f))=\Gamma(ufu^{-1})$. Hence, the transformation of the local
coordinate system is of the form
\begin{equation}\label{coordinates_transformation}
\begin{array}{c}
\widetilde{u}:Hom(X,X^\perp)\To Hom(u(X),u(X^\perp)) \\
\\
\widetilde{u}(f)=ufu^{-1}
\end{array}
\end{equation}
\\
\bigskip
The tangent space of the manifold $\gr{m}{\hilb}$ at a point $X\in\gr{m}{\hilb}$ can be identified with
the vector space $Hom(X,X^\perp)$ (see \cite{Milnor-Stasheff}). The identification of the manifold \gr{m}{\hilb} with the
manifold of projections $\proj_m(\hilb)$, gives another representation of the tangent space
$T_X(\gr{m}{\hilb})$, which follows from \ref{projections}:
\begin{equation}\label{projections_tangent}
\begin{array}{l}
T_X(\gr{m}{\hilb})\cong\{\Phi:\hilb\To\hilb\,|\,\Phi\textrm{ is linear and }\\
\proj(X)\circ\Phi+\Phi\circ\proj(X)=\Phi,\,\Phi^*=\Phi,\,\trace{\Phi}=0\}
\end{array}
\end{equation}
This s exactly the set of such operators $\Phi:\hilb\To\hilb$, that the decomposition of $\Phi$,
corresponding to the decomposition of the Hilbert space $\hilb=X\oplus X^\perp$ is of the form
$$
\Phi=
\left[
\begin{array}{cc}
0       & \varphi^* \\
\varphi & 0
\end{array}
\right]
$$
The differential of the mapping $f\mapsto\proj(\Gamma(f))$ (see \ref{projection_in_coordinates}),
at the point $0\in Hom(X,X^\perp)$, gives the isomorphism between the two representations of the
tangent space $T_X(\gr{m}{\hilb})$:
\begin{equation}\label{tangent_isomorphism}
Hom(X,X^\perp)\ni\varphi\mapsto\proj'_X(\varphi)=
\left[
\begin{array}{cc}
0       & \varphi^* \\
\varphi & 0
\end{array}
\right]
\end{equation}
\\
\bigskip
Recall that an action of a Lie group $G$ on a smooth manifold $M$, gives rise of the
homomorphism from the Lie algebra of $G$ to the Lie algebra of vector fields on the manifold $M$:
for $v\in\mathfrak{g}$, where $\mathfrak{g}$ denotes the Lie algebra of the Lie group $G$, the
corresponding vector field $\widetilde{v}$, induced by the action of the group on $M$, is defined
as (see \cite{Nomizu})
$$
\widetilde{v}(x)=\frac{\partial F}{\partial G}(1,x)(v)
$$
where $F:G\times M\To M$ is the mapping corresponding to the action of the group $G$ on the
manifold $M$.

So, the action of the group $U(\hilb)$ on the manifold \gr{m}{\hilb}, gives rise of the
homomorphism from the Lie algebra of $U(\hilb)$, to the Lie algebra of vector fields on the
manifold \gr{m}{\hilb}. To describe this homomorphism, recall that if $\proj(X)$ is the
orthogonal projection corresponding to the element $X\in\gr{m}{\hilb}$, then for any orthogonal
transformation $g\in U(\hilb)$, the orthogonal projection corresponding to $g(X)$ is
$\proj(g(X))=g\proj(X)g^{-1}$. Therefore, for any fixed point $X\in\gr{m}{\hilb}$, we have a
mapping
$$
U(\hilb)\ni g\mapsto g\proj(X)g^{-1}\in\proj_m(\hilb)\cong\gr{m}{\hilb}
$$
This implies that the tangent vector corresponding to an element $u\in\mathfrak{u}(\hilb)$ of the
Lie algebra of the Lie group $U(\hilb)$, at a point $X\in\gr{m}{\hilb}$ is
$\widetilde{u}_X=[u,\proj(X)]$, where $[\cdot\,,\,\cdot]$ denotes the commutator of two
operators. If the decomposition of the linear operator $u:\hilb\To\hilb$, corresponding to the
decomposition $\hilb=X\oplus X^\perp$ is
$$
u=
\left[
\begin{array}{cc}
U_{XX} & -U^*_{XX^\perp} \\
& \\
U_{XX^\perp} & U_{X^\perp X^\perp}
\end{array}
\right]
$$
then it is clear that
$$
\widetilde{u}_X=[u,\proj(X)]=
\left[
\begin{array}{cc}
0 & U^*_{XX^\perp} \\
&\\
U_{XX^\perp} & 0
\end{array}
\right]
$$
This, together with the formula \ref{tangent_isomorphism}, implies that the vector field
corresponding to the Lie algebra element $u\in\mathfrak{u(\hilb)}$ via the action of the group
$U(\hilb)$ on $\gr{m}{\hilb}$ is
\begin{equation}\label{lie_algebra_tangent}
\widetilde{u}_X=U_{XX^\perp}=(1-\proj(X))\circ u\in Hom(X,X^\perp)\cong T_X(\gr{m}{\hilb})
\end{equation}
\section[Some Operations Over Connections]
{Some Operations Over Connections\\on Vector Bundles}
In this section we shall review some facts from the theory of vector bundles and connections in
the context of this work.

For any vector bundle $\pi:E\To M$, we denote by $S(E)$ the space of smooth sections of this
vector bundle.

Let $V\stackrel{p}{\To}M$ and $W\stackrel{q}{\To}M$ be two vector bundles over a smooth manifold $M$. Let
$S(V)\stackrel{\ssup{\nabla}{V}}{\To}S(T^*(M)\otimes V)$ and
$S(W)\stackrel{\ssup{\nabla}{W}}{\To}S(T^*(M)\otimes W)$ be connections (covariant derivations)
on them. We define the following operations over these connections:\\
\textbf{The direct (or Whitney) sum of two connections.} Define a connection
$\nabla=\ssup{\nabla}{V}\oplus\ssup{\nabla}{W}$ on the Whitney sum of the vector bundles
$(V,M,p)$ and $(W,M,q)$ as follows: for a vector field $\xi$ on the manifold $M$ and a section
$s=s_1\oplus s_2$ of the vector bundle $V\oplus W$ let
\begin{equation}\label{connections_sum}
\nabla_\xi(s)=\ssup{\nabla}{V}_\xi(s_1)+\ssup{\nabla}{W}_\xi(s_2)
\end{equation}
\\
\textbf{The tensor product of two connections.} Define a connection
$\nabla=\ssup{\nabla}{V}\otimes\ssup{\nabla}{W}$ on the tensor product of the vector bundles
$(V,M,p)$ and $(W,M,q)$ as
\begin{equation}\label{connections_product}
\nabla_\xi(s_1\otimes s_2)=\ssup{\nabla}{V}_\xi(s_1)\otimes s_2+
s_1\otimes\ssup{\nabla}{W}_\xi(s_2)
\end{equation}
\\
\textbf{The connection on the dual vector bundle.} Define a connection
$\ssup{\nabla}{{V^*}}$ on the vector bundle $p_*:V^*\To M$ a fiber of which at a point
$x\in M$ is the dual vector space of the vector space $p^{-1}(x)$
\begin{equation}\label{connection_dual}
\xi\scalar{s}{\tau}=\scalar{\ssup{\nabla}{V}_\xi(s)}{\tau}+
\scalar{s}{\ssup{\nabla}{{V^*}}_\xi(\tau)}
\end{equation}
where $s$ is a section of the vector bundle $V$ and $\tau$ is a section of the dual vector bundle
$V^*$, and $\scalar{s}{\tau}$ is the function on the manifold $M$ obtained by the pairing of the
sections $s$ and $\tau$.\\
\textbf{The connection on the bundle of homomorphisms.} Let
$$
Hom(V,W)\stackrel{h}{\To}M
$$
be the fiber bundle the fiber of which at a point $x\in M$ is the space of all homomorphisms
from the vector space $p^{-1}(x)$ to the vector space $q^{-1}(x)$. In the case of finite
dimensional fibers, the fiber bundle is canonically isomorphic to the fiber bundle
$h:V^*\otimes W\To M$. We have defined the connections on the dual spaces bundle and the
tensor products bundle, therefore, the covariant derivation
$\nabla=Hom(\ssup{\nabla}{V},\ssup{\nabla}{W})=\ssup{\nabla}{{V^*}}\otimes\ssup{\nabla}{W}$
on the space of sections $S(Hom(V,V))$ can be defined as $\nabla_\xi(\tau\otimes s)=
\ssup{\nabla}{{V^*}}_\xi(\tau)\otimes s+\tau\otimes\ssup{\nabla}{W}_\xi(s)$ for $\tau\in S(V^*)$
and $s\in S(W)$. This implies that for $t\in S(V)$ we have the following
$$
\begin{array}{l}
\nabla_\xi(\tau\otimes s)(t)=\scalar{\ssup{\nabla}{{V^*}}_\xi(\tau)}{t}\cdot s+
\tau(t)\cdot\ssup{\nabla}{W}_\xi(s)=\\
\\
=\xi(\tau(t))\cdot s+\tau(t)\cdot\ssup{\nabla}{W}_\xi(s)-
\scalar{\tau}{\ssup{\nabla}{V}_\xi(t)}\cdot s=\\
\\
=\ssup{\nabla}{W}_\xi(\tau(t)\cdot s)-\scalar{\tau}{\ssup{\nabla}{V}_\xi(t)}\cdot s
\end{array}
$$
which suggests the following more elegant expression for the covariant derivation
$\nabla=Hom(\ssup{\nabla}{V},\ssup{\nabla}{W})$ on the space of sections $S(Hom(V,W))$:
\begin{equation}\label{connections_hom}
\begin{array}{c}
\textrm{ for }f\in S(Hom(V,W))\textrm{ and }t\in S(V)\textrm{ let}\\
\\
\nabla_\xi(f)(t)=\ssup{\nabla}{W}_\xi(f(t))-f(\ssup{\nabla}{V}_\xi(t))
\end{array}
\end{equation}
It is easy to verify that the results of all the above defined operations satisfy the conditions
required for covariant derivations on vector bundles.

Let $p:V\To M$ and $q:W\To M$ be such vector bundles over a smooth manifold $M$ that their
Whitney sum is a trivial vector bundle: $V\oplus W\cong M\times H$. This situation gives rise of
a connection on the vector bundles $(V,M,p)$ and $(W,M,q)$ as follows. Any section of the vector
bundle $V$ can be regarded as a function on $M$ with values in the vector space $H$. Let us
denote this function under $\widetilde{s}:M\To H$. For a vector field $\xi$ on the manifold $M$,
define the covariant derivation of the section $s$ as
$\nabla_\xi(s)(x)=P_{V_x}(\widetilde{s}'_x(\xi_x)),\;\forall x\in M$. Here $P_{V_x}$ is the
projection operator $P_{V_x}:H\To V_x$ corresponding to the decomposition $H=V_x\oplus W_x$. The
covariant derivation on the sections of the vector bundle $W$ can be defined analogically. It
can be verified directly that the above defined operation satisfies the conditions required for
covariant derivations.
\begin{rem}\label{remark1}
If $V_1,V_2,W_1$ and $W_2$ are vector bundles over the manifold $M$ such that
$V_1\oplus W_1\cong M\times H_1$ and $V_2\oplus W_2\cong M\times H_2$ and $\nabla^1$ and
$\nabla^2$ are the connections on $V_1$ and $V_2$, accordingly, as defined above, then the
covariant derivation on the vector bundle $V_1\otimes V_2$ induced by the decomposition
$$
\Big(V_1\otimes V_2\Big)\oplus
\Big((V_1\otimes W_2)\oplus(W_1\otimes V_2)\oplus(W_1\otimes W_2)\Big)\cong
M\times\Big(H_1\otimes H_2\Big)
$$
is the same as $\nabla=\nabla^1\otimes\nabla^2$ (see the formula \ref{connections_product}).
\end{rem}
\begin{defn}
Let $\pi:T(M)\To M$ be the tangent vector bundle for a smooth manifold $M$ and $p:V\To M$ be
such vector bundle that the Whitney sum $T(M)\oplus V$ is trivial bundle $M\times H$. We call the
decomposition of a trivial fiber bundle, $M\times H=T(M)\oplus V$, \textbf{integrable} if the induced covariant derivation
$\nabla$ on the sections of the tangent bundle (i.e., vector fields on $M$) satisfies the
condition
\begin{equation}\label{integrable_connection}
\nabla_X(Y)-\nabla_Y(X)=[X,Y]
\end{equation}
for any two vector fields $X$ and $Y$ on the manifold $M$.
\end{defn}
If the vector space $H$, which is the fiber of the Whitney sum, is a Hilbert space, and for each
point $x\in M$, the vector space $V_x$ (the fiber of the vector bundle $V$ at the point $x$) is
the orthogonal complement of the subspace $T_x(M)\subset H$, then the covariant derivation
$\nabla$ satisfies the condition
$$
X\scalar{Y}{Z}=\scalar{\nabla_X(Y)}{Z}+\scalar{Y}{\nabla_X(Z)}
$$
Hence, we can state that the connection $\nabla$, induced by the decomposition
$M\times H=T(M)\oplus V$, is the covariant derivation corresponding to the Levi-Civita connection
on the manifold $M$, for the metric induced by the embedding $T(M)\subset M\times H$.
\begin{rem}\label{connection_on_tangent}
If $\Phi:M\To H$ is a smooth mapping, such that for each point $x\in M$ the linear mapping
$\Phi'_x:T_x(M)\To H$ is a monomorphism, then the vector bundle $p:\Phi^\perp(H)\To M$, where
$p^{-1}(x)=\image{\Phi'_x}^\perp$, is such that the Whitney sum $T(M)\oplus\Phi^\perp(H)$ is
trivial: $M\times H$; and the connection (covariant derivation) induced by the decomposition
$M\times H=T(M)\oplus\Phi^\perp(H)$ is the Levi-Civita connection on $T(M)$.
\end{rem}

Let $\pi:P\To M$ be a principal bundle with a structure group $G$, acting from right on the
total space $P$. Let $F$ be a vector space on which the Lie group $G$ acts from left. The
associated vector bundle over the manifold $M$ is defined as the vector bundle with the total
space $(P\times V)/G\equiv P_V$, where the quotient is taken under the right action of the group
$G$ on the manifold $P\times V$: $(p,v)\mapsto (pg, g^{-1}v)$, for each $(p,v)\in P\times G$ and
$g\in G$ (see \cite{Nomizu}). The projection mapping $\pi^V:P_V\To M$ for this vector bundle is defined as
$$
\pi^V([p,v])=\pi(p),\quad\forall\;[p,v]\in(P\times V)/G
$$
It follows from the definition that any section of the associated vector bundle can be regarded
as a fuction $\varphi:P\To V$, such that
$\varphi(pg)=g^{-1}\varphi(p),\;\forall p\in P,\;\forall g\in G$. Let us denote the space of such
functions on the total space of the principal bundle $P$, by $C^\infty(P,V)_G$.

A connection form $A$ on the principle bundle $(P,M,\pi)$ is a differential 1-form, with values
in $\mathfrak{g}$ -- the Lie algebra of the Lie group $G$, such that for any $g\in G$:
$R^*_g(A)=Ad(g^{-1})(A)$; and for any $u\in\mathfrak{g}$: $A(\widetilde{u})=u$; where
$\widetilde{u}$ denotes the tangent vector
$$
\widetilde{u}\in T_p(P),\quad\widetilde{u}=\frac{d}{dt}(p\cdot\exp(t\cdot u))
$$
The connection form $A$ defines a distribution of horizontal subspaces on the total space $P$:
$$
H_p=\kernel{A_p}\subset T_p(P),\quad\forall\;p\in P
$$
This distribution is carried to the total space of the associated vector bundle, by the quotient
mapping: $P\times V\To P_V=(P\times V)/G$.

A vector field $X$ on the manifold $M$ can be lifted to a ``horizontal'' vector field
$\widetilde{X}$ on the total space $P$, which is uniquely defined by the following conditions
$$
\widetilde{X}_p\in\kernel{A_p},\;\forall\;p\in P\textrm{ and }\pi'(\widetilde{X})=X
$$
As it was mentioned, a section of the associated vector bundle $s\in\Gamma(P_V)$ can be identified with
an element $\varphi_s\in C^\infty(P,V)_G$. The covariant derivation of the section $s$ is the
section of the vector bundle $(P_V,M,\pi_V)$, corresponding to the function
$\varphi_s'(\widetilde{X})\in C^\infty(P,V)_G$.
\section[The Geometry of The Canonical Vector Bundle]
{The Geometry of The Canonical Vector\\Bundle on Grassmann Manifold}
Consider the set $\can{m}{\hilb}=\set{(X,x)\,|\,X\in\gr{m}{\hilb},\,x\in X}$, which is the
subset of $\gr{m}{\hilb}\times\hilb$. The set $\can{m}{\hilb}$ together with the differential
structure induced from $\gr{m}{\hilb}\times\hilb$ is a differential manifold. The mapping
$$
p:\can{m}{\hilb}\To\gr{m}{\hilb},\quad p(X,x)=X
$$
is a vector bundle over the Grassmann manifold $\gr{m}{\hilb}$. The fiber at a point
$X\in\gr{m}{\hilb}$ is the vector space $X\subset\hilb$, itself. This vector bundle is known as
the \emph{canonical vector bundle} over the Grassmann manifold $\gr{m}{\hilb}$. For any point
$X\in\gr{m}{\hilb}$, the local trivialization of this vector bundle, in the neighborhood of the
point $X$, naturally arises from the coordinate system:
$$
\Phi_X:Hom(X,X^\perp)\times X\To\can{m}{\hilb},\quad\Phi_X(\varphi,x)=
(\Gamma(\varphi),x+\varphi(x))
$$
If we regard the manifold $\gr{m}{\hilb}$ as the manifold of orthonormal projectors,
$\proj_m(\hilb)$ (see \ref{projections}), we can describe the total space $\can{m}{\hilb}$, as a
submanifold of $\proj_m(\hilb)\times\hilb$ defined by the equation
$P(x)=x,\;(P,x)\in\proj_m(\hilb)\times\hilb$. The latter implies that the tangent space of the
manifold $\can{m}{\hilb}$ is defined by the equation $dP(x)+P(dx)=dx$. More precisely: for any
point $Q=(P,x)\in\can{m}{\hilb}$, where $P\in\proj_m(\hilb)\cong\gr{m}{\hilb}$ is a projector
and $x\in\image{P}=X$, the tangent space $T_Q(\can{m}{\hilb})$ is the subspace of the space
$Hom(X,X^\perp)\times\hilb$ defined by the linear equation:
$\varphi(x)=(1-P)u,\;\varphi\in Hom(X,X^\perp),\;u\in\hilb$. That is
\begin{equation}\label{canonical_tangent}
T_{(X,x)}(\can{m}{\hilb})=\set{(\varphi,u)\in Hom(X,X^\perp)\times\hilb\;|\;\varphi(x)=\proj(X^\perp)(u)}
\end{equation}
From this description of the tangent space of the total space of the canonical fiber bundle,
easily follows that the vertical tangent subspace at a point $(X,x)\in\can{m}{\hilb}$ is
$$
Vert(X,x)=\set{(0,u)\in Hom(X,X^\perp)\times\hilb\,|\,u\in X}
$$
One natural choice of a horizontal complement of the vertical subspace could be
$$
Hor(X,x)=\set{(\varphi,\varphi(x))\,|\,\varphi\in Hom(X,X^\perp)}
$$
The projection operator on the vertical tangent subspace, along the horizontal tangent subspace
is
\begin{equation}\label{vertical_projection}
P_{Vert}:T_{(X,x)}(\can{m}{\hilb})\To Vert(X,x),\quad P_{Vert}(\varphi,w)=(0,w-\varphi(x))
\end{equation}
\begin{prop}\label{can_covariant}
The distribution of horizontal subspaces
$$
Hor(X,x)\subset T_{(X,x)}(\can{m}{\hilb}),\;(X,x)\in\can{m}{\hilb}
$$
is a connection on the canonical vector bundle $\can{m}{\hilb}$. The corresponding covariant
derivation acts on the space of sections of $\can{m}{\hilb}$ as follows: any section
$s\in S(\can{m}{\hilb})$ can be regarded as a function $s:\gr{m}{\hilb}\To\hilb$; for any tangent
vector $\varphi\in T_X(\gr{m}{\hilb})\cong Hom(X,X^\perp)$, the covariant derivative of $s$ by
$\varphi$ is $\nabla_\varphi(s)=\proj(X)(s_X'(\varphi))=s_X'(\varphi)-\varphi(s(X))$.
\end{prop}
\begin{prf}
As it follows from the formula for the projection on the vertical tangent subspace
(see \ref{vertical_projection}), the action of the differential of the function $s$ on the
tangent vector $\varphi$ and then its projection on the vertical tangent space is
$s_X'(\varphi)-\varphi(s(X))$. As it follows from the definition of the tangent space of
$\can{m}{\hilb}$ (see \ref{canonical_tangent}), it is the same as the projection of
$s_X'(\varphi)$ on the subspace $X$ along its orthogonal complement $X^\perp$.\\
It can be verified by direct calculations that the operation
$\nabla_\varphi(s)=S'(\varphi)-\varphi\circ s$ has the properties of covariant derivation.
\end{prf}
Now consider the vector bundle $q:\canp{m}{\hilb}\To M$ the fiber of which at a point
$X\in\gr{m}{\hilb}$ is $X^\perp$ -- the orthogonal complement of $X$. It is clear that the
Whitney sum of two vector bundles $\can{m}{\hilb}\oplus\canp{m}{\hilb}$ is the trivial vector
bundle: $\gr{m}{\hilb}\times\hilb$. As in the case of $\can{m}{\hilb}$, we can describe the
total space $\canp{m}{\hilb}$ as a submanifold in $\proj_m(\hilb)\cong\gr{m}{\hilb}\times\hilb$:
$$
\canp{m}{\hilb}=\set{(P,y)\in\proj_m(\hilb)\times\hilb\,|\,P(y)=0}
$$
Differentiating the equation $P(y)=0$, we obtain the equation for the tangent space of
$\canp{m}{\hilb}$: $dP(y)+P(dy)=0$. This implies that the tangent space of $\canp{m}{\hilb}$ at a
point $(X,y),\;X\in\gr{m}{\hilb},\;y\in X^\perp$ is the subspace of $Hom(X,X^\perp)\times\hilb$:
\begin{equation}\label{canonical_perp_tangent}
T_{(X,y)}(\canp{m}{\hilb})=\set{(\varphi,u)\in Hom(X,X^\perp)\times\hilb\,|\,\varphi^*(y)+\proj(X)(u)=0}
\end{equation}
From this description of the tangent space of $\canp{m}{\hilb}$, follows that the vertical
tangent subspace for the fiber bundle $q:\canp{m}{\hilb}\To M$ at a point $(X,y)$ is
\begin{equation}\label{perp_vertical_tangent}
Vert(X,y)=\set{(0,u)\in Hom(X,X^\perp)\times\hilb\,|\,u\in X^\perp}
\end{equation}
One choice of the complementar (``horizontal'') subspace is
$$
Hor(X,y)=\set{(\varphi,-\varphi^*(y))\in Hom(X,X^\perp)\times\hilb}
$$
The discussion analogical to that in the proof of the proposition \ref{can_covariant}, shows
that the covariant derivation corresponding to this distribution of horizontal subspaces is
$$
\nabla^\perp_\varphi(s)=s'_X(\varphi)+\varphi^*(s(X))=\proj(X^\perp)(s'_X(\varphi)))
$$
for $\varphi\in Hom(X,X^\perp)\cong T_X(\gr{m}{\hilb})$ and $s$ is a section of the vector bundle
$\canp{m}{\hilb}$ (in this case regarded as a function $s:\gr{m}{\hilb}\To\hilb$).

It is clear that $\can{m}{\hilb}\oplus\canp{m}{\hilb}=M\times\hilb$. Using the definition of the
covariant derivation on the sum of two vector bundles (see the formula \ref{connections_sum}), we
obtain a covariant derivation on the trivial vector bundle $M\times\hilb$:
$\ssup{\nabla}{\hilb}=\nabla\oplus\nabla^\perp$
\begin{equation}\label{sum_covariant_derivation}
\ssup{\nabla}{\hilb}_\varphi(s)=s'_X(\varphi)-\varphi(s_1(X))+\varphi^*(s_2(X))
\end{equation}
where $X\in\gr{m}{\hilb},\,\varphi\in Hom(X,X^\perp)\cong T_X\gr{m}{\hilb}$, $s_1$ is a section
of $\can{m}{\hilb}$, $s_2$ is a section of $\canp{m}{\hilb}$ and $s=s_1+s_2$ is a section of
$\gr{m}{\hilb}\times\hilb$ (that is, a function $\gr{m}{\hilb}\To\hilb$).

The formula \ref{sum_covariant_derivation} for the covariant derivation on $\gr{m}{\hilb}\times\hilb$
implies that the corresponding connection form $\mathcal{F}$ is a 1-form on the manifold
$\gr{m}{\hilb}$ with values in the Lie algebra of the group of unitary transformations of the
Hilbert space $\hilb$ (which is the Lie algebra antisymmetric operators on $\hilb$):
$$
\begin{array}{c}
\mathcal{F}_X(\varphi)=
\left[
\begin{array}{ccc}
0 & \ & \varphi \\
 & &             \\
-\varphi^* & \ & 0
\end{array}
\right]\in\mathfrak{u}(\hilb)\\
\\
\textrm{for }X\in\gr{m}{\hilb}\textrm{ and }\varphi\in T_X(\gr{m}{\hilb})\cong Hom(X,X^\perp)
\end{array}
$$
As it was mentioned (see the formula \ref{tangent_isomorphism}), the projective representation of
the Grassmann manifold gives rise of the 1-form with values in the space of symmetric operators
$$
d\proj(\varphi)=
\left[
\begin{array}{ccc}
0 & \, & \varphi^* \\
& & \\
\varphi & \, & 0
\end{array}
\right],\;\varphi\in T_X(\gr{m}{\hilb})\cong Hom(X,X^\perp)
$$
Using this form, the connection form $\mathcal{F}$, can be written as a 1-form on
$\proj_m(\hilb)\cong\gr{m}{\hilb}$, as
\begin{equation}\label{connection_form_on_grassmann}
\mathcal{F}=\proj d\proj+(\proj-1)d\proj=2\proj d\proj-d\proj
\end{equation}
Its differential (the curvature form) is $d\mathcal{F}=2d\proj d\proj$, which is a 2-form with
values in the Lie algebra of antisymmetric operators on the Hilbert space $\hilb$
$$
(d\mathcal{F})(\varphi,\psi)=2
\left[
\begin{array}{ccc}
\varphi^*\psi-\psi^*\varphi & \, & 0 \\
&&\\
0 & \, & \varphi\psi^*-\psi\varphi^*
\end{array}
\right],\quad\varphi,\psi\in Hom(X,X^\perp)
$$
\\
\bigskip
\newcommand{\selfadj}[1]{\mathfrak{S}(#1)}
Let us denote by $\selfadj{\hilb}$ the space of symmetric operators on the Hilbert space $\hilb$.
The empedding $\proj:\gr{m}{\hilb}\To\selfadj{\hilb}$ of the Grassmann manifold as the set of
projectors, induces a mapping $\proj':T(\gr{m}{\hilb})\To\selfadj{\hilb}\times\selfadj{\hilb}$
$$
\proj'_X(\varphi)=(\proj(X),\widetilde{\varphi})
$$
where
$$
\widetilde{\varphi}=
\left[
\begin{array}{cc}
0       & \varphi^* \\
\varphi & 0
\end{array}
\right]
$$
This mapping, according to the remark \ref{connection_on_tangent} induces a connection on the
tangent bundle $T(\gr{m}{\hilb})$, which itself, coincides with the connection
$\nabla^*~\otimes~\nabla^\perp$, where $\nabla$ is the connection on the canonical vector bundle
$\can{m}{\hilb}$ and $\nabla^\perp$ is the connection on $\canp{m}{\hilb}$.
\section[Grassmann Manifold as an Orbit of the (co-)Adjoint Action]
{Grassmann Manifold as an Orbit of the (co-)Adjoint Action and the Corresponding Symplectic Structure}
Let $L$ be a finite-dimensional Lie algebra over the field of complex or real numbers and $L^*$
be its dual vector space. The linear mapping
$$
B:L~\wedge~L~\To L,\quad B(x\wedge y)=[x,y]
$$
induces the dual mapping between the dual spaces
$$
B^*:L^*\To L^*\wedge L^*,\quad B^*(\varphi)(x\wedge y)=\varphi([x,y])
$$
The latter can be regarded as an antisymmetric covariant tensor field on the linear space $L^*$.
Such a field defines a bracket on the algebra of smooth functions $C^\infty(L^*)$:
$$
\set{f,g}(\varphi)=(df|_\varphi\wedge dg|_\varphi)(B^*(\varphi))=
\varphi([df|_\varphi,dg|_\varphi])
$$
where the elements of the space $L^{**}$: $df|_\varphi$ and $dg|_\varphi$ are considered as
elements of the vector space $L$ which is canonicaly isomorphic to $L^{**}$. The algebra
$C^\infty(L^*)$ together with the above defined bracket is a Poisson algebra. Particularly, if
we take any two elements $x,y\in L$, they can be regarded as linear functions on the vector space
$L^*$: $x(\varphi)=\varphi(x),\;\forall\;\varphi\in L^*$; and the Poisson bracket of these two
linear functions is $\set{x,y}=[x,y]$.

If the vector space $L$ is equipped with a scalar product $\scalar{\cdot\;}{\cdot}$, we can carry the
Poisson structure from the space $L^*$ to $L$ by the linear mapping
$u\mapsto\scalar{u}{\cdot}:L\To L^*$. Under these conditions, an element $x\in L$ can be
considered as a linear function on the vector space $L$:
$\widetilde{x}(u)=\scalar{x}{u},\;\forall\,u\in L$. The Poisson bracket of such two functions is
$$
\set{\widetilde{x},\widetilde{y}}(u)=\scalar{[x,y]}{u}=\widetilde{[x,y]}(u),\;\forall\,u\in L
$$
Now, assume that $L$ is a Lie algebra of a Lie group $G$ and the scalar product on the vector
space $L$ is invariant under the adjoint action of the group $G$:
\begin{equation}\label{adjoint_invariant}
\scalar{gug^{-1}}{gvg^{-1}}=\scalar{u}{v},\quad\forall\,u,v\in L,\;\forall\,g\in G
\end{equation}
Suppose that $g=\exp(tw),\;w\in L,t\in\mathbb{R}$ is a one-parameter subgroup generated by an
element $w$. Differentiating the equality \ref{adjoint_invariant} by the parameter $t$ at the
point $t=0$, we obtain the following infinitezimal version of the equality
\ref{adjoint_invariant}
\begin{equation}\label{adjoint_invariant_infinitezimal}
\scalar{[w,u]}{v}+\scalar{u}{[w,v]}=0
\end{equation}
\newcommand{\ham}[1]{\textrm{ham}(#1)}
For any function $f\in C^\infty(L)$, we denote by $\ham{f}$ the Hamiltonian vector field
corresponding to the function $f$ for the Poisson structure defined by $\scalar{\;}{}$. For any
element $x\in L$ and the corresponding linear function $\widetilde{x}=\scalar{x\,}{\cdot}$ on
$L$, we have  the following
$$
\begin{array}{c}
d\widetilde{y}(\ham{\widetilde{x}})=\set{\widetilde{x},\widetilde{y}}=\widetilde{[x,y]}\gamomd\\
\gamomd\;\forall\,u\in L:\widetilde{y}(\ham{\widetilde{x}_u})=\widetilde{[x,y]}(u)=\scalar{[x,y]}{u}
\end{array}
$$
Keeping in mind the equality \ref{adjoint_invariant_infinitezimal}, we obtain the following
$$
\ham{\widetilde{x}}_u=[x,u],\;\forall\,u\in L
$$
That is: for a Poisson structure defined by an adjoint-invariant metric on a Lie algebra $L$, the
Hamiltonian vector field corresponding to a linear function
$\widetilde{x}=\scalar{x}{\cdot},\;x\in L$ is $\ham{\widetilde{x}}_u=[x,u],\;\forall u\in L$.
Hence, we obtain that for any point $u\in L$, the subspace of the tangent space $T(L)\cong L$,
generated by the Hamiltonian vector fields at the point $u$ is
$$
Ham_u=\set{[x,u]\,|\,x\in L}
$$
The distribution of the subspaces of the tangent spaces: $u\mapsto Ham_u,\,u\in L$, is
integrable and the integral submanifolds of this distribution are the orbits of the adjoint action
of the corresponding Lie group $G$. These orbits are exactly the symplectic leaves of the Poisson
structure and as it should be, the restriction of the Poisson structure on each of them is
nondegenerated. The corresponding symplectic form on any orbit of the adjoint action is
\begin{equation}\label{orbit_symplectic_form}
\omega(x;u,v)=\scalar{x}{[u,v]},\;x,u,v\in L
\end{equation}
\\
\bigskip
Any curve $u:\mathbb{R}\To L$ can be regarded as a time-dependent Hamiltonian:
$\widetilde{u}_t(x)=\scalar{u(t)}{x},\;t\in\mathbb{R},x\in L$; and the corresponding Hamiltonian
equation is $\dot{x}(t)=[u(t),x(t)]$. As the symplectic leaves of the Poisson structure are the
orbits of the adjoint action, we have tha tany solution of the Hamiltonian equation lies in some
orbit of the adjoint action.

\bigskip
Now consider the situation when the Lie algebra $L$ is the Lie algebra of antisymmetric operators
on the Hilbert space $\hilb$. This Lie algebra is the same as the Lie algebra of the Lie group $U(\hilb)$ --
the group of unitary transformations of the Hilbert space $\hilb$. We denote this Lie algebra
by $\mathfrak{u}(\hilb)$. The mapping $\proj_m(\hilb)\To\mathfrak{u}(\hilb),\quad P\mapsto -iP$
is an injection of the Grassmann manifold in the Lie algebra $\mathfrak{u}(\hilb)$. The equality
$\proj(g(X))=g\proj(X)g^{-1},\;\forall\,X\in\gr{m}{\hilb},\;\forall\,g\in U(\hilb)$, implies that
the image of the above mapping -- $i\proj_m(\hilb)$, is an orbit of the adjoint action of the Lie
group $U(\hilb)$ on its Lie algebra -- $\mathfrak{u}(\hilb)$. Therefore, the Grassmann manifold
can be regarded as an orbit of an adjoint action of a Lie group on its Lie algebra.

Consider the following scalar product on the vector space $\mathfrak{u}(\hilb)$:
$$
\scalar{u}{v}=\trace{u^*\circ v},\quad u,v\in\mathfrak{u}(\hilb)
$$
For any unitary transformation $g\in U(\hilb)$, we have
$$
\scalar{gug^{-1}}{gvg^{-1}}=\trace{gu^*vg^{-1}}=\trace{u^*v}=\scalar{u}{v},\;\forall\,u,v\in\mathfrak{u}(\hilb)
$$
which implies that the scalar product on the space $\mathfrak{u}(\hilb)$ is invariant under the
adjoint action of the group $U(\hilb)$. Hence, we have the symplectic form on any orbit of the
adjoint action, defined by the formula \ref{orbit_symplectic_form}. Particularly, for the
embedded Grassmann manifold in $\mathfrak{u}(\hilb)$, we have
\begin{equation}\label{symplectic_on_grassmann1}
\omega(iP;\Phi,\Psi)=\trace{iP\circ[\Phi,\Psi]}
\end{equation}
To obtain more explicit expression for the differential 2-form $\omega$, at a point
$P\in\proj_m(\hilb)\cong\gr{m}{\hilb}$, consider the decomposition of the operators $\Phi$ and
$\Psi$ corresponding to the decomposition $\hilb=\image{P}\oplus\image{P}^\perp$:
$$
\Phi=
\left[
\begin{array}{ccc}
0 & \, & i\varphi^* \\
&&\\
i\varphi & & 0
\end{array}
\right],\qquad
\Psi=
\left[
\begin{array}{ccc}
0 & \, & i\psi^* \\
&&\\
i\psi & & 0
\end{array}
\right]
$$
where $\varphi,\psi:\image{P}\To\image{P}^\perp$. From the formula \ref{symplectic_on_grassmann1}
easily follows the following expression for the symplectic form:
\begin{equation}\label{symplectic_on_grassmann2}
\begin{array}{l}
\omega(X;\varphi,\psi)=i\cdot\trace{\varphi^*\psi+\psi^*\varphi)}\\
\\
\textrm{for }X\in\gr{m}{\hilb}\textrm{ and }\varphi,\psi\in T_X(\gr{m}{\hilb})\cong Hom(X,X^\perp)
\end{array}
\end{equation}
which can be written as $w=i\trace{P\cdot dP\wedge dP}$.

As it was mentioned for the case of a general Lie algebra, any element $u\in\mathfrak{u}(\hilb)$
can be considered as a function $\widetilde{u}:\gr{m}{\hilb}\To\mathbb{R}$
$$
\widetilde{u}(P)=i\cdot\trace{u^*\circ P}=-i\cdot\trace{u\circ P},\quad
P\in\proj_m(\hilb)\cong\gr{m}{\hilb}
$$
and the corresponding Hamiltonian vector field on the manifold \gr{m}{\hilb} is
\begin{equation}\label{ham_field}
\ham{\widetilde{u}}_P=[u,P],\;P\in\proj_m(\hilb)\cong\gr{m}{\hilb}
\end{equation}
This vector field is the same as that one generated by the action of the group $U(\hilb)$.
\newcommand{\isom}{Isom(\mathbb{C}^m,\hilb)}
\section[The Connection on the Canonical Principal Bundle]
{The Canonical Principal Bundle on Grassmann Manifold and the Canonical Connection}
Consider the set of all orthonormal m-frames in the Hilbert space $\hilb$. This set is known as
the Stiefel manifold (see \cite{Milnor-Stasheff}) for the complex Hilbert space $\hilb$. It is clear that the Stiefel
manifold of orthonormal m-frames can be identified with the set of all isometric mappings from
the complex vector space $\mathbb{C}$ to the Hilbert space $\hilb$. We denote this set of all
isometric mappings from $\mathbb{C}$ to $\hilb$ by $\isom$. This set equipped with the topology
and the differential structure of the subset of the vector space $Hom(\mathbb{C}^m,\hilb)$, is a
manifold, and can be described as follows
$$
\isom=\set{\varphi\in Hom(\mathbb{C}^m,\hilb)\,|\,\varphi^*\varphi=1}
$$
There is a natural mapping from the manifold $\isom$ to the manifold $\gr{m}{\hilb}$ defined as
$$
\pi:\isom\To\gr{m}{\hilb},\qquad\pi(\varphi)=\image{\varphi}
$$
Recall that for any $X\in\gr{m}{\hilb}$, the symbol $\proj(X)$ denotes the operator on the Hilbert
space $\hilb$ that is the orthogonal projection on the subspace $X$.
\begin{lem}
For any $\varphi\in\isom$, we have that $\proj(\image{\varphi})=\varphi\circ\varphi^*$ (see \cite{Fujii-1}, \cite{Fujii-2}, \cite{Fujii-3})
\end{lem}
\begin{prf}
The proof consists of the following two steps: firstly we check that for any $x\in\image{\varphi}$:
$(\varphi\varphi^*)(x)=x$, and then we check that for any $y\in\image{\varphi}^\perp$:
$(\varphi\varphi^*)(y)=0$.

For $x\in\image{\varphi}$, we have the following:
$$
x=\varphi(u)\;\gamomd\;(\varphi\varphi^*)(x)=(\varphi\underbrace{\varphi^*\varphi}_1)(u)=
\varphi(u)=x
$$
For $y\in\image{\varphi}^\perp$, we have:
$$
\begin{array}{c}
\forall\,u\in\hilb:\;\scalar{(\varphi\varphi^*)(y)}{u}=\scalar{\varphi^*(y)}{\varphi^*(u)}=
\scalar{y}{\underbrace{(\varphi\varphi^*)(y)}_{\in\image{\varphi}}}=0\;\gamomd\\
\gamomd\;(\varphi\varphi^*)(y)=0
\end{array}
$$
\end{prf}
This implies that that, as the manifold $\gr{m}{\hilb}$ is identified with the space of
projectors $\proj_m(\hilb)$, in some cases, we can substitute the mapping
$$
\pi:\isom\To\gr{m}{\hilb}
$$
with the mapping
$$
\pi:\isom\To\proj_m(\hilb),\qquad\varphi\mapsto\varphi\varphi^*
$$
The latter we also denote by $\pi$.

For any $X\in\gr{m}{\hilb}$, consider any isometric map $\varphi:\mathbb{C}^m\To X$. It is clear
that $\pi(\varphi)=X$. Hence the mapping $\pi$ is surjective and for each $X\in\gr{m}{\hilb}$ we
have that
$$
\pi^{-1}(X)=\set{\varphi\in\isom\,|\,\image{\varphi}=X}
$$
\begin{lem}
Let $\set{u_1,\cdots,u_m}$ be any basis of a Hermitian vector space $X$. There exists one and only one
orthogonal basis $\set{e_1,\cdots,e_m}$ in the space $X$, such that for any $i=1,\cdots,m$:
$Span<e_1,\cdots,e_i>=Span<u_1,\cdots,u_i>$ and
$e_1\wedge\cdots\wedge e_i=k\cdot u_1\wedge\cdots\wedge u_i$, where $k>0$ (i.e., the orientations
of the frames $\set{e_1,\cdots,e_i}$ and $\set{u_1,\cdots,u_i}$ coincide).
\end{lem}
\begin{prf}
We prove this lemma by induction. For $m=1$, it is clear that the basis $e_1$ should be
$u_1/\norm{u_1}$. If the statement is true for $m-1$ then consider a vector of the type
$e'=x_1e_1+\cdots+x_{m-1}e_{m-1}+u_m$. The system of equations
$\scalar{e'}{e_i}=0,\;i=1,\ldots,m-1$ gives the values of the numbers $x_1,\ldots,x_{m-1}$. After
this, we can obtain the vector $e_m$ by normalizing the vector $e'$: $e_m=c\cdot e'$; so that
$\norm{e_m}=1$ and the orientation of the frame $\set{e_1,\ldots,e_m}$ be the same as the
orientation of $\set{u_1,\ldots,u_m}$ and it is clear that this could be done in exactly one way.
\end{prf}
The statement of the above lemma implies that for any isomorphism $f:\mathbb{C}^m\To X$, there is
exactly one linear transformation $I(f):X\To X$, such that $\Phi(f)=I(f)\circ f:\mathbb{C}^m\To X$
is an isometric mapping and
$Span<\Phi(f)(e_1),\ldots,\Phi(f)(e_i)>=Span<f(e_1),\ldots,f(e_i)>$ and the orientations of the
frames $\set{\Phi(f)(e_1),\ldots,\Phi(f)(e_i)}$ and $\set{f(e_1),\ldots,f(e_i)}$ coincide, for
$i=1,\ldots,m$, where $e_1,\ldots,e_m$ is the natural basis in the complex vector space
$\mathbb{C}^m$. We call the mapping $\Phi(f)$ the \emph{isometrization} of the isomorphism $f$.

By using of the above construction we can define a local trivialization of
$\pi:\isom\To\gr{m}{\hilb}$ as follows: for any point $X\in\gr{m}{\hilb}$, consider the mapping
$$
\begin{array}{ll}
\Phi_X:Isom(\mathbb{C}^m,X)\times Hom(X,X^\perp)\To\isom\\
\\
\Phi_X(\varphi,f)=\Phi((1+f)\varphi)
\end{array}
$$
It is clear that the mapping $(1+f)\varphi$ is a monomorphism, the image of which is the subspace
$\Gamma(f)\in\gr{m}{\hilb}$, and $\Phi((1+f)\varphi)$ is the isometrization of the mapping
$(1+f)\varphi$.

Hence, we obtain that $\pi:\isom\To\gr{m}{\hilb}$ is a locally trivial fiber bundle.

Consider the following right action of $U(m)$ -- the group of unitary transformations of
$\mathbb{C}^m$, on the space $\isom$:
$$
\forall\,g\in U(m)\textrm{ and }\forall\,\varphi\in\isom\textrm{ let }\varphi\mapsto\varphi\circ g
$$
It is clear that $\pi(\varphi g)=\pi(\varphi)$.

\newcommand{\complex}[1]{\mathbb{C}^{#1}}

If $\varphi$ and $\psi$ are in one and the same fiber $\pi^{-1}(X)$ for $X\in\gr{m}{\hilb}$,
then $\varphi\varphi^*=\psi\psi^*$, which implies that
$\varphi=\psi\underbrace{\psi^*\varphi}_g=\psi g$. For $g=\psi^*\varphi$ we have
$g^*g=\varphi^*\underbrace{\psi\psi^*}_1\varphi=\varphi^*\varphi=1$, which implies that
$g$ is the element of the unitary group $U(m)$. We obtain that the right action of the group
$U(m)$ on the fiber of the bundle $\pi:\isom\To\gr{m}{\hilb}$ is transitive and effective. In
other words, this fiber bundle is a principal bundle with the structure group $U(m)$. The
associated vector bundle $(\isom\times\mathbb{C}^m)/U(m)$ where
$$
(\varphi,x)\sim(\varphi\circ g,g^{-1}(x)),
\;\forall\,(\varphi,x)\in\isom\times\mathbb{C}^m\textrm{ and }\forall\,g\in U(m)
$$
is isomorphic to the canonical vector bundle $\can{m}{\hilb}$ via the mapping
$$
[\varphi,x]\mapsto\varphi(x)
$$
The subset $\isom$ in the space $Hom(\complex{m},\hilb)$ is defined by the equation
$\varphi^*\varphi=1$, which implies that the tangent space of $\isom$ can be described by the
equation
\begin{equation}\label{tangent_of_principal1}
d\varphi^*\varphi+\varphi^*d\varphi=0
\end{equation}
Otherwise
\begin{equation}\label{tangent_of_principal2}
T_\varphi(\isom)=\set{u\in Hom(\complex{m},\hilb)\,|\,u^*\varphi+\varphi^*u=0}
\end{equation}
\\
\bigskip
Differentiating the mapping $\pi(\varphi)=\varphi\varphi^*$ from $\isom$ to \gr{m}{\hilb}, we
obtain
\begin{equation}\label{pi_diff}
d\pi=d\varphi\varphi^*+\varphi d\varphi^*
\end{equation}
\begin{prop}
The vertical tangent subspace of the fiber bundle
$$
\pi:\isom\To\gr{m}{\hilb}
$$
at a point $\varphi\in\isom$ is the subspace of the tangent space $T_\varphi$ consisting of such
$v:\complex{m}\To\hilb$ that $\image{v}\subset\image{\varphi}$.
\end{prop}
\begin{prf}
By definition, the vertical tangent space at a point $\varphi\in\isom$ is the kernel of the
mapping $\pi'(\varphi)$. Hence, it is described by the following system of linear equations
$$
\begin{cases}
v^*\varphi+\varphi^*v=0 &  \textrm{-- Tangent}\\
v\varphi^*+\varphi v^*=0 & \textrm{-- Vertical}
\end{cases}
$$
The multiplication of the second one by $\varphi$ from the right, gives $v+\varphi v^*\varphi=0$.
Substitute the term $v^*\varphi$ by $-\varphi^*v$, from the first equation, gives the following
equation $v-\varphi\varphi^*v=0$, or equivalently $(1-\proj(\image{\varphi}))v=0$. But it is clear
that the operator $(1-\proj(\image{\varphi}))$ is the orthogonal projection operator on the
subspace $\image{\varphi}^\perp$. This implies that $\image{v}\subset\image{\varphi}$.
\end{prf}
For a given subspace $X\subset\hilb$, any linear mapping $w:\complex{m}\To\hilb$ is uniquely
decomposed as a direct sum: $w=v+u$, with $\complex{m}\stackrel{v}{\To}X$ and
$\complex{m}\stackrel{u}{\To}X^\perp$. The above description of the vertical tangent space for the
fiber at a point $X\in\gr{m}{\hilb}$ suggests the following natural choice of its complementar
(horizontal) subspace
$$
Hor_X=\set{u\in Hom(\complex{m},\hilb)\,|\,\image{u}\subset X^\perp}
$$
For any $\varphi\in\isom$ with $\image{\varphi}=X$, we have that $\kernel{\varphi}=X^\perp$,
therefore: $\varphi^*u=0$, which implies that $(\varphi^*u)^*=0$ and finally:
$u^*\varphi+\varphi^*u=0$. Therefore, the space $Hor_X$ automatically is subspace of the tangent
space $T_\varphi(\isom)$.

\newcommand{\connform}{\mathcal{A}}
\newcommand{\connformx}{\varphi^*d\varphi}

The distribution of the horizontal subspaces $\varphi\mapsto Hor_\varphi$ can be described by the
equation $\varphi^*d\varphi=0$. Consider the following differential form
$\connform=\varphi^*d\varphi$ on the manifold $\isom$.
\begin{prop}
The differential form $\connform=\connformx$ takes its values in the Lie algebra of the Lie
group $U(m)$, and is a connection form on the total space of the $U(m)$-principal bundle
$\pi:\isom\To\gr{m}{\hilb}$.
\end{prop}
\begin{prf}
From the equation for the tangent space of $\isom$: $\varphi^*d\varphi+d\varphi^*\varphi=0$,
follows that the differential form $\connform$ takes its values in the Lie algebra of
antisymmetric matrices: $\connform^*=d\varphi^*\varphi=-\varphi^*d\varphi=-\connform$.

The next step is to verify the transformation rule for the differential form $\connform$ under
the right action of the unitary group $U(m)$. For any $g\in U(m)$, we have the following
$$
R_g(\varphi)=\varphi g\;\gamomd\;R_g^*(\connform)=g^*\varphi^*d(\varphi g)=
g^{-1}\varphi^*d\varphi g=g^{-1}\connform g=Ad(g^{-1})(\connform)
$$
which shows that the transformation rule for the differential form $\connform$ is in accordance
with the requirement for the connection forms.

For any fixed point $\varphi\in\isom$, consider the mapping
$$
m_\varphi:U(m)\To\isom,\quad m_\varphi(g)=\varphi g,\;\forall\,g\in U(m)
$$
Its differential at the point $g=1$ is the linear mapping from the Lie algebra $\mathfrak{u}(m)$
to the tangent space $T_\varphi(\isom)$:
$m'_\varphi(1)(u)=\varphi u,\;\forall\,u\in\mathfrak{u}(m)$. Consider the value of the differential form
$\connform$ on the vertical tangent vector $m'_\varphi(1)(u)=\varphi u$. We obtain
$\connform(\varphi u)=\underbrace{\varphi^*\varphi}_1u=u$. The latter eqality is also in accordance
with the requirement for the connection forms.
\end{prf}
Let $\xi\in T_\varphi(\isom)$ be a horizontal tangent vector. That is:
$\xi\in Hom(\complex{m},\hilb)$ and $\image{\xi}\subset\image{\varphi}^\perp$. Its image in
$T_{\pi(\varphi)}\gr{m}{\hilb}$ by the differential of the projection map $\pi$ is
$$
\pi'(\varphi)(\xi)=\xi\varphi^*+\varphi\xi^*=
\left[
\begin{array}{cc}
0 & \varphi\xi^* \\
\xi\varphi^* & 0
\end{array}
\right]:\image{\varphi}\oplus\image{\varphi}^\perp\To\image{\varphi}\oplus\image{\varphi}^\perp
$$
Otherwise, we can consider $\xi$ as a linear mapping from $\complex{m}$ to $\image{\varphi}^\perp$,
$\varphi$ as a linear mapping from $\complex{m}$ to $\image{\varphi}$, and $\pi'(\varphi)(\xi)$, as
the composition
$$
\xi\circ\varphi^*:\image{\varphi}\stackrel{\varphi^*}{\To}\complex{m}\stackrel{\xi}{\To}\image{\varphi}^\perp
$$
which is the element of
$Hom(\image{\varphi},\image{\varphi}^\perp)\cong T_{\image{\varphi}}(\gr{m}{\hilb})$. Vice versa,
any tangent vector $\mu\in Hom(X,X^\perp)$ at a point $X\in\gr{m}{\hilb}$, could be lifted to the
corresponding horizontal space at a point $\varphi\in\isom$ as $\widetilde{mu}=\mu\varphi$, which
is a mapping from $\complex{m}$ to $X^\perp$ (i.e., the element of the horizontal subspace in
$T_\varphi(\isom)$). It is clear that
$\widetilde{\xi\varphi^*}=\xi\underbrace{\varphi^*\varphi}_1=\xi$.
\section[the Holonomy Algebra of the Connection Form $\varphi^*d\varphi$]
{Geometric Control Theory and the Holonomy Algebra of the Connection Form $\varphi^*d\varphi$}
Let us recall the following well known result from the theory of connections and their holonomies
on principal bundles.
\begin{thm}\textbf{(Ambrous, Singer).}
Let $A$ be a connection form on a principle bundle $\pi:P\To M$. For any point $p\in P$, the
holonomy algebra $\Phi_p$ at the point $p$ is the algebra spanned on the set of elements of the
type $\Omega(X,Y)$, where $\Omega$ is the curvature form for the connection form $A$ and $X$ and
$Y$ are horizontal tangent vectors of $P$, at such points $q$ that can be connected with the
point $p$ by a horizontal curve.
\end{thm}\label{holonomy_algebra_thm}
We use the statement of this theorem to investigate the holonomy algebra of the connection form
$\connform=\varphi^*d\varphi$ on the canonical principal bundle over the Grassmann manifold.
\begin{prop}
For any point $\varphi\in\isom$, the holonomy algebra of the connection
$\connform=\varphi^*d\varphi$ at this point coincides with the Lie algebra of the Lie group
$U(\hilb)$.
\end{prop}
\begin{prf}
As it was shown, the horizontal subspace corresponding to the connection form $\connform$, at a
point $\varphi\in\isom$ is
$$
Hor(\varphi)=\set{u\in Hom(\complex{m},\hilb)\,|\,\image{u}\subset\image{\varphi}^\perp}
$$
and the value of the curvature form on a pair of horizontal vectors $(u,v)$ at the point
$\varphi$ is
$$
\Omega(u,v)=(d\connform)(u,v)=\frac{1}{2}(u^*v-v^*u)
$$
We show that for any antisymmetric operator $w:\complex{m}\To\complex{m}$, can be found such a
pair of linear maps $u,v:\complex{m}\To\image{\varphi}^\perp$ that $w=\frac{1}{2}(u^*v-v^*u)$.
In this case we assume that the Hilbert space is infinite-dimensional (or, at least has a
dimension as big as we need). So, we can choose an m-dimensional complex subspace in
$\image{\varphi}^\perp$. Fix any basis in $X$, after which it is identified with $\complex{m}$.
If we take $u=1:\complex{m}\To X\cong\complex{m}$, then from the equation
$w=\frac{1}{2}(v-v^*)$, easily follows the solution $v=w$.
\end{prf}
Hence, for the Grassmann manifold of an infinite-dimensional Hilbert space, it is not necessary
to use the full strength of the theorem \ref{holonomy_algebra_thm}, because the $\image{\Omega}$,
only in one point of the total space $\isom$, covers all the Lie algebra $\mathfrak{u}(m)$,
even without spanning. The same is true for a finite-dimensional case, but when
$\dim(\image{\varphi}^\perp)\geq m$. In the case when the Hilbert space is finite-dimensional and
$\dim(\image{\varphi}^\perp)<m$, things are not so simple.
\begin{prop}
For any integer $n>m$ and any $\psi\in Isom(\complex{m},\complex{n})$, the holonomy algebra of
the connection form $\connform=\varphi^*d\varphi$, at the point $\psi$, coincides with the entire
Lie algebra $\mathfrak{u}(m)$.
\end{prop}
\begin{prf}
Let us assume, that the antisymetric operator $w:\complex{m}\To\complex{m}$ is diagonal
$$
w=
\left(
\begin{array}{cccc}
a_1   & 0      & \cdots & 0     \\
\cdot & \cdot  & \cdots & \cdot \\
\cdot & \cdot  & \cdots & \cdot \\
0     & 0      & \cdots & a_m
\end{array}
\right),\quad\textrm{ where }a_i\in\mathbf{i}\cdot\mathbb{R},\,i=1,\ldots,m
$$
It is sufficient to consider the case when $n=m+1$. We have that $w=\sum\limits_{i=1}^mw_i$, where
$$
w_i=
\left(
\begin{array}{cccccc}
0 & \cdot & \cdot & \cdot & \cdot & 0 \\ 
0 & \cdot & \cdot & \cdot & \cdot & 0 \\ 
0 & \cdot & a_i   & \cdot & \cdot & 0 \\ 
0 & \cdot & \cdot & \cdot & \cdot & 0 \\ 
0 & \cdot & \cdot & \cdot & \cdot & 0 \\ 
0 & \cdot & \cdot & \cdot & \cdot & 0
\end{array}
\right),\qquad i=1,\cdots,m
$$
For each $w_i$ consider the pair of linear mappings $u_i,v_i:\complex{m}\To\complex{}$, where
$u_i=(\underbrace{0,\cdots,0}_{i-1},1,0,\cdots,0)$ and
$v_i=(\underbrace{0,\cdots,0}_{i-1},a_i,0,\cdots,0)$. For these mappings we have that
$w_i=\frac{1}{2}(u_i^*v_i-v_i^*u_i)$. Hence, we obtain that any diagonal matrix $w$ can be
represented as $w=\sum\limits_i\Omega(u_i,v_i)$, where $u_i,v_i:\complex{m}\To\complex{}$. Now,
recall that for any antisymmetric operator $w:\complex{m}\To\complex{m}$ there exists a unitary
transformation $\tau:\complex{m}\To\complex{m}$, such that $\tau w\tau^{-1}$ is diagonal. If
$\tau w\tau^{-1}=\sum\limits_i\Omega(u_i,v_i)$, then consider the operators
$\widetilde{u_i}=u_i\tau$ and $\widetilde{v_i}=v_i\tau$. We have that
$w=\sum\limits_i\Omega(\widetilde{u_i},\widetilde{v_i})$.
\end{prf}
Consider the holonomy of the connection $\connform=\varphi^*d\varphi$ from the point of view of
the geometric control theory. First of all, let us recall some definitions and facts from the
geometric control theory (see \cite{Lorby}).
\begin{defn}\textbf{(Control Group).}
Let $I$ be any non-empty set. Consider the set of finite sequences of the type
$((t_1,i_1)(t_2,i_2),\ldots,(t_p,i_p))$ with entries from the set $\mathbb{R}\times I$. Introduce
the following reduction rules
\begin{itemize}
\item
In any such sequence, the entries of the type $(0,k)$ are removed;
\item
If $i_k=i_{k+1}$, then the segment $(t_k,i_k)(t_{k+1},i_{k+1})$ is replaced by the item
$(t_k+t_{k+1},i_k)$.
\end{itemize}
Applying the above reduction procedures, to any sequence, we obtain a non-reducible sequence,
after a finite number of steps. The set of non-reducible sequences of the elements of the set
$\mathbb{R}\times I$ is called the \textbf{control set} corresponding to the set $I$. We denote
the set of such sequences by $C(I)$. The elements of the set $C(I)$ are called the
\textbf{controls}.
\end{defn}
From any two elements of the control set $C(I)$, $s_1$ and $s_2$, we can construct a new one by
concatenation $s_1s_2$ and then by reduction of the result -- $Red(s_1s_2)$. The mapping from
the set $C(I)\times C(I)$ to $C(I)$:
$$
(s_1,s_2)\mapsto Red(s_1s_2)
$$
defines a group structure on the control set $C(I)$. The uital element in this group is the
\textbf{empty} sequence. Further, the product of the elements $s_1$ and $s_2$ we denote by
$s_1s_2$ or $s_1\cdot s_2$.

For any control $s=((t_1,i_1),\cdots,(t_n,i_n))\in C(I)$ and a real number $a\in\mathbb{R}$,
define the control $a\cdot s$ as
$$
a\cdot s=((at_1,i_1),\cdots,(at_n,i_n))
$$
Any set of controls $\set{s_1,s_2,\ldots,s_p}\subset C(I)$, defines a mapping
$$
\phi_{s_1s_2\ldots s_p}:\mathbb{R}^p\To C(I)
$$
as
$$
(a_1,\ldots,a_p)\mapsto (a_1\cdot s_1,\ldots,a_p\cdot s_p),\quad\forall\,
(a_1,\ldots,a_p)\in\mathbb{R}^p
$$
Define the topology on the set $C(I)$ as the strongest topology for which, all the mappings
$\set{\phi_{s_1\cdots s_p}\,|\,p\in\mathbb{N},\;s_i\in C(I),\;i=1,\ldots,p}$ are continuous.

The differential structure on the topological space $C(I)$ is defined as follows: any map
$f:C(I)\To\mathbb{R}$ is differentiable if and only if all the maps
$f\circ\phi_{s_1\cdots s_p}:\mathbb{R}^p\To\mathbb{R}$, are differentiable.
\begin{defn}
A \textbf{dynamical polysystem} on a smooth manifold $M$, controlled by the group of control $C(I)$
is called a smooth left action of the group $C(I)$ on the manifold $M$. For any point $x\in M$,
the set $C(I)x=\set{sx\,|\,s\in C(I)}$ is called the orbit of the point $x$.
\end{defn}
Any element $i\in I$ defines a one parameter group of diffeomorphisms
$$
\varphi^i_t:M\To M,\quad\varphi^i_t(x)=(t,i)x
$$
which, itself, generates a vector field $X^i$ on the manifold $M$. The family of vector fields
$\set{X_i\,|\,i\in I}$ is called the \emph{infinitezimal transformations} of the dynamical
polysystem $(M,C(I))$.

Conversely, for any family of vector fields $\set{X_i\,|\,i\in I}$ on a smooth manifold $M$, the
formula
$$
(t_1,i_1)\cdots(t_1,i_1)x=(\varphi_{t_1}^{i_1}\circ\cdots\circ\varphi_{t_n}^{i_n})(x)
$$
where $\varphi_{t_1}^{i_1},\ldots\varphi_{t_n}^{i_n}$ are the elements of the one-parameter groups
of diffeomorphisms corresponding to the vector fields $X_{i_1},\dots,X_{i_n}$, defines a dynamical
polysystem controlled by the group $C(I)$.

To summarize, we can say that there is a one-to-one correspondence between the dynamical
polysystems and the families of vector fields.

Further, for any subset of vector fields $X$ on a smooth manifold $M$, the dynamical polysystem
controlled by the group $C(X)$ will be denoted also by $C(X)$.

One of the purposes of the geometric control theory is the investigation of the accessibility
problem: for a given dynamical polysystem $C(X)$, corresponding to some family of vector fields
$X$, find the orbit of a point $m\in M$ --- $C(X)m$. Let us formulate the following two
fundamental theorems about the structure of the orbit of a dyanamical polysystem, corresponding
to a family of vector fields.
\begin{thm}\textbf{(Orbits Theorem. Nagano-Sussmann).}
Let $X$ be a family of vector fields on a smooth manifold $M$ and $m$ is a point on $M$. Then:
\begin{enumerate}
\item
The orbit $C(X)m$ is an immersed submanifold in the manifold $M$;
\item
The tangent space $T_x(C(X)m)$ at a point $x\in C(X)m$ is the vector space generated by the set of vectors
$\set{Ad(s)(u_x)\,|\,s\in C(X),\,u\in X}$.
\end{enumerate}
\end{thm}
For any family of vector fields $X$, let us denote by $Lie(X)$ the minimal submodule of
$C^\infty(M)$-module of vector fields on the manifold $M$, containing the family $X$ and closed
under the operation of Lie bracket.
\begin{defn}
A family of vector fields $X$ is called \textbf{completely nonholonomic} or
\textbf{bracket-generating} if for each point $m\in M$ we have that $Lie(X)_m=T_m(M)$.
\end{defn}
The next fundamental theorem of the geometric control theory is, practically, a corollary of the
previous one.
\begin{thm}\textbf{(Rashevsky-Chow).}
Let $M$ be a connected smooth manifold, and let $X$ be a family of vector fields on the manifold
$M$. If the family $X$ is completely nonholonomic (i.e., $Lie(X)_m=T_m(M),\,\forall\,m\in M$)
then the orbit $C(X)m$ coincides with the entire manifold $M$, for each point $m\in M$.
\end{thm}

\newcommand{\lietr}[1]{\mathfrak{u}(#1)}
\newcommand{\imm}{\hookrightarrow}

Let $V$ be a complex subspace of the Hilbert space $\hilb$. As before, $U(\hilb)$ is the group of
unitary transformations of the Hilbert space $\hilb$ and $\mathfrak{u}(V)$ is the Lie algebra of
the Lie group of unitary transformations of the subspace $V$. Consider the following
$\lietr{V}$-valued 1-form on the Lie group $U(\hilb)$:
$$
\mathcal{B}=\imath^*(g^*dg)\imath
$$
where $\imath$ denotes the natural immersion mapping $\imath:V\imm\hilb$ and $\imath^*$ is its
dual $\imath^*:\hilb\To V$. The value of the form $\mathcal{B}$ can be interpreted as the
$(g^{-1}dg)_{VV}$ component of the opertor $g^{-1}dg\in\lietr{\hilb}$ in the decomposition
$$
g^{-1}dg=
\left[
\begin{array}{ccc}
(g^{-1}dg)_{VV}       &&      (g^{-1}dg)_{V^\perp V} \\
                      &&                             \\
(g^{-1}dg)_{VV^\perp} && (g^{-1}dg)_{V^\perp V^\perp}
\end{array}
\right]
$$
corresponding to the decomposition of the Hilbert space $\hilb=V\oplus V^\perp$. The differential
form $\mathcal{B}$ is left-invariant: for any $a\in U(\hilb)$ we have the following
$$
L_a^*(\mathcal{B})=\imath^*(g^*a^*adg)\imath=\imath^*(g^*dg)\imath=\mathcal{B}
$$

Consider the distribution of tangent subspaces on the manifold $U(\hilb)$ defined by the equation
$\mathcal{B}=0$, i.e., the subspace of the tangent space at a point $u\in U(\hilb)$ is the kernel
of the 1-form $\mathcal{B}$ at the point $u$. As the differential form $\mathcal{B}$ is
left-invariant, the distribution $X$ is left-invariant too: $X_{gu}=L'_g(X_u)$. Therefore, the
distribution $X$ is of a constant rank. But $X$ is not an involutive distribution. It follows
from the definition of the form $\mathcal{B}$: the subset of the tangent spce
$X_{1}\in T_1(U(\hilb))$, consists of the operators of the form
$$
f=
\left[
\begin{array}{ccc}
0       & & -\varphi^* \\
        & &            \\
\varphi & & 0
\end{array}
\right]:\hilb=V\oplus V^\perp\To\hilb=V\oplus V^\perp
$$
If we have two such operators
$$
f_1=
\left[
\begin{array}{ccc}
0       &  & -\varphi_1^* \\
        &  &            \\
\varphi_1  & & 0
\end{array}
\right]\textrm{ and }
f_2=
\left[
\begin{array}{ccc}
0       &  & -\varphi_2^* \\
        &  &            \\
\varphi_2  & & 0
\end{array}
\right]
$$
and $u_1,u_2\in X$ are the left-invariant vector fields on $U(\hilb)$, generated by $f_1$ and
$f_2$, respectively, we have that
$$
[u_1,u_2]_1=[f_1,f_2]=
\left[
\begin{array}{ccc}
\varphi_2^*\varphi_1-\varphi_1^*\varphi_2 &  & 0                                         \\
                                          &  &                                           \\
0                                         &  & \varphi_2\varphi_1^*-\varphi_1\varphi_2^*
\end{array}
\right]
$$
Let us denote the family of vector fields defined by the distribution $X$, also by $X$ and
suppose that the subspce $V$ in the Hilbert spce $\hilb$ is finite-dimensional.
\begin{prop}
The family of vector fields $X$ on the manifold $U(\hilb)$ is completely nonholonomic.
\end{prop}
\begin{prf}
Let $u_0$ and $u_1$ be any two elements of $U(\hilb)$. We have to show that there exists such
control $(t_1,X_1)\cdots(t_n,X_n),\,t_i\in\mathbb{R},\,X_i\in X$, that
$\exp(t_1X_1)\cdots\exp(t_nX_n)u_0=u_1$.

As the distribution $X$ is left-invariant, we can assume that $u_0=1$. Consider the fiber bundle
$\pi:Isom(V,\hilb)\To\gr{m}{\hilb}$, where $m=\dim(V)$, and the connection form
$\connform=\varphi^*d\varphi$ on it. As it follows from the theorem about the holonomy algebra
of the connection $\connform$, any two points $\varphi_0,\varphi_1\in Isom(V,\hilb)$, can be
connected by a piece-wise smooth curve
$$
\Phi:[0,1]\To Isom(V,\hilb),\qquad\Phi(0)=\varphi_0,\,\Phi(1)=\varphi_1
$$
Assume that $\varphi_0$ and $\varphi_1$ are such that $\varphi_0(v)=v$ and
$\varphi_1(v)=u_1(v),\,\forall\,v\in V$. Consider the fiber bundle
$\pi^\perp:Isom(V^\perp,\hilb)\To\gr{m}{\hilb}$, the fiber of which at a point
$W\in\gr{m}{\hilb}$ is $Isom(V^\perp,W^\perp)$. Consider two points
$\varphi_0^\perp,\varphi_1^\perp\in Isom(V^\perp,\hilb)$, such that $\varphi_0^\perp(x)=x$ and
$\varphi_1^\perp(x)=u_1(x),\,\forall\,x\in V^\perp$. It is clear that
$\varphi_0\oplus\varphi_0^\perp=\mathbf{1}$ and $\varphi_1\oplus\varphi_1^\perp=u_1$. On the
manifold $\gr{m}{\hilb}$ we have a curve $\alpha:[0,1]\To\gr{m}{\hilb},\,\alpha=\pi\Phi$, for
which $\alpha(0)=V$ and $\alpha(1)=u_1(V)$. Let $\Phi^\perp:[0,1]\To Isom(V^\perp,\hilb)$ be such
lifting of the curve $\alpha$ that $\Phi^\perp(0)=\varphi_0^\perp$ and
$\Phi^\perp(1)=\varphi_1^\perp$. It is clear that $\set{u(t)=(\Phi+\Phi^\perp)(t)\,|\,t\in[0,1]}$
is a piece-wise smooth family of unitary transformations of the Hilbert space $\hilb$, such that
$u(t)\varphi_0=\Phi(t),\,u(0)=1,\,u(1)=u_1,\;t\in[0,1]$. As the curve $\Phi$ is horisontal, we
obtain that $\varphi_0^*u^*\dot{u}\varphi_0=0$. Hence, we obtain that the piece-wise smooth curve
$u(t),\,t\in[0,1]$, connects the points $u_0=1$ and $u_1$ in $U(\hilb)$, and at the same time is
an integral curve of the dynamical polysystem $X$ on the Lie group $U(\hilb)$.
\end{prf}.
\section[ Schr\"{o}dinger Equation and Berry Phase]
{The Action of the Group $U(\hilb)$ on the Principal Bundle $\isom$. Schr\"{o}dinger\\
Equation and Berry Phase}
Any element $g$ of the unitary group $U(\hilb)$ acts on the total space $\isom$ as
$\varphi\mapsto g\varphi$, and on the base $\gr{m}{\hilb}$ as $X\mapsto g(X)$ (or, on the
language of projectors: $P\mapsto gPg^{-1}$). These two actions are correlated:
$$
\pi(g\varphi)=g\pi(\varphi),\,\forall\,\varphi\in\isom\textrm{ and }\forall\,g\in U(\hilb)
$$
It other words, the unitary transformation of the Hilbert space $\hilb$ defines an automorphism
of the principal bundle $\pi:\isom\To\gr{m}{\hilb}$. The connection form
$\connform=\varphi^*d\varphi$ is invariant under the action of such automorpism
$$
(g\varphi)^*d(g\varphi)=\varphi^*g^{-1}gd\varphi=\varphi^*d\varphi
$$
For any element $u$ of the Lie algebra $\mathfrak{u}(\hilb)$, we have a one-parameter subgroup of
the group $U(\hilb)$generated by $u$ --- $\set{\exp(tu),\;t\in\mathbb{R}}$. The action of this
group on the total space $\isom$ defines a vector field $\widehat{u}$, which can be described as
$\widehat{u}_\varphi=u\circ\varphi,\;\forall\,\varphi\in\isom$. The action of this group on the
base $\gr{m}{\hilb}$, defines a vector field
$\check{u}_P=[u,P],\;\forall\,P\in\proj_m(\hilb)\cong\gr{m}{\hilb}$ which is the Hamiltonian
vector field corresponding to the function $\widetilde{u}:\gr{m}{\hilb}\To\mathbb{R}$ defined as
$\widetilde{u}(P)=\mathbf{i}\cdot\trace{u^*\circ P}$ (see \ref{ham_field}). The tangent vector
$\check{u}$ at a point $X\in\gr{m}{\hilb}$, can be described as $\check{u}_X=u_{XX^\perp}$, where
$u_{XX^\perp}\in Hom(X,X^\perp)$ is a component in the decomposition
$$
u=
\left[
\begin{array}{ccc}
u_{XX}       & & -u_{XX^\perp}^*     \\
             & &                     \\
u_{XX^\perp} & &  u_{X^\perp X^\perp}
\end{array}
\right]:\hilb=X\oplus X^\perp\To\hilb=X\oplus X^\perp
$$
As the actions of the group $U(\hilb)$ on the total space and the base are compatible, we have
that $\pi'(\hat{u})=\check{u}$.
\begin{rem}
This can be checked directly:
$$
\begin{array}{l}
\pi(\varphi)=\varphi\varphi^*\;\gamomd\;d\pi=d\varphi\varphi^*+\varphi d\varphi^*\;\gamomd\\
\\
\gamomd\;d\pi(\hat{u})=u\varphi\varphi^*+\varphi(u\varphi)^*=
u\underbrace{\varphi\,\varphi^*}_{\pi(\varphi)}-\underbrace{\varphi\,\varphi^*}_{\pi(\varphi)}u=
[u,\pi(\varphi)]
\end{array}
$$
\end{rem}
As the action of the group $U(\hilb)$ preserves the connection form $\connform$, we have that
$L_{\hat{u}}(\connform)=0$, where $L_{\hat{u}}$ denotes the Lie derivation by the vector field
$\hat{u}$.
\begin{rem}
It can be checked directly:
$$
\begin{array}{l}
L_{\hat{u}}(\connform)=d(\connform(\hat{u}))+(d\connform)(\hat{u},\cdot)=\\
\\
=d(\varphi^*u\varphi)+(u\varphi)^*d\varphi-d\varphi^*(u\varphi)=
d\varphi^*u\varphi+\varphi^*ud\varphi-\varphi^*ud\varphi-d\varphi^*u\varphi=0
\end{array}
$$
\end{rem}
The vertical component of the vector field $\hat{u}$ is measured by the connection form $\connform$
and is
\begin{equation}\label{vert_component}
\varphi\connform(\hat{u})=\varphi\varphi^*u\varphi=\pi(\varphi)u\varphi
\end{equation}
Therefore, the horizontal component of the vector field $\hat{u}$ at the point $\varphi\in\isom$
is
\begin{equation}\label{hor_component}
u\varphi-\pi(\varphi)u\varphi=(1-\pi(\varphi))u\varphi
\end{equation}
It is clear that in the expression \ref{vert_component}, the term $\pi(\varphi)$ is the projection
operator on thesubspace $\image{\varphi}\subset\hilb$ and the term $1-\pi(\varphi)$, in the
expression \ref{hor_component}, is the projection operator on the subspace
$\image{\varphi}^\perp\subset\hilb$. Let us describe the vertical and horizontal components of the
tangent vector $u\varphi$ more explicitly. For any point $\varphi\in\gr{m}{\hilb}$, let
$$
u=
\left[
\begin{array}{ccc}
u_{11} & & -u_{21}^* \\
&&\\
u_{21} & & u_{22}
\end{array}
\right]
$$
be the decomposition of the operator $u\in U(\hilb)$, corresponding to the decomposition of the
Hilbert space $\hilb=\image{\varphi}\oplus\image{\varphi}^\perp$. From the formulas \ref{vert_component}
and \ref{hor_component}, follows that the vertical component of the vector $u\circ\varphi$ is
$$
u_{11}\circ\varphi:\complex{m}\To\image{\varphi}
$$
and the horizontal component is
$$
u_{21}\circ\varphi:\complex{m}\To\image{\varphi}^\perp
$$
\\
\bigskip
Let $H(t),\,t\in [0,T],\,T>0$ be a time-dependent Hamiltonian, i.e., for any $t\in [0,T]$, $H(t)$
is an antisymmetric operator on the Hilbert space $\hilb$. This can be regarded as a time-dependent
$C^\infty$-class function (Hamiltonian)
$$
\tilde{H(t)}:\gr{m}{\hilb}\cong\proj_m(\hilb)\To\mathbb{R},\qquad
\tilde{H(t)}(P)=\mathbf{i}\cdot\trace{H(t)^*\circ P}
$$
which, together with the symplectic form (see the formula \ref{symplectic_on_grassmann2})
$\omega(X;\varphi,\psi)=i\cdot\trace{\varphi^*\psi+\psi^*\varphi)}$, defines a time-dependent
Hamiltonian vector field on the Grassmann manifold $\gr{m}{\hilb}$:
$ham(\tilde{H(t)})_P=[H(t),P],\,\forall\,P\in\proj_m(\hilb)$ (or equivalently:
$ham(\tilde{H(t)})_X=H(t)_{XX^\perp}\in Hom(X,X^\perp),\forall\,X\in\gr{m}{\hilb}$), and the
corresponding time-dependent vector field $\check{H(t)}$ on the total space $\isom$:
$\check{H(t)}_\varphi=H(t)\circ\varphi,\forall\,\varphi\in\isom$. The corresponding differential
equation on the space $\isom$ is
\begin{equation}\label{lifted_schrodinger}
\dot{\varphi}(t)=H(t)\circ\varphi(t),\,t\in[0,T]
\end{equation}
and the differential equation (Hamiltonian (or Schr\"{o}dinger) equation) on the Grassmann
manifold is
\begin{equation}\label{schrodinger}
\dot{P}(t)=[H(t),P(t)],\,t\in[0,T]
\end{equation}
Let $P(t),\;t\in[0,T],\;P(0)\equiv P_0$ be a solution of the equation \ref{schrodinger} and
$\varphi(t),\;t\in[0,T],\;\varphi(0)\cong\varphi_0\in\pi^{-1}(P_0)$ be the corresponding solution
of the equation \ref{lifted_schrodinger}, in the total space $\isom$. It is clear that
$\pi(\varphi(t))=P(t),\,\forall\,t\in[0,T]$, but the curve $\varphi(t)$, in general, is not the
horizontal lifting of the curve $P(t)$, defined by the connection form $\connform=\varphi^*d\varphi$,
because, the tangent vector $\dot{\varphi}(t)$ has the vertical component --- $H(t)_{11}\circ\varphi(t)$,
which, in general, is different from zero. This, pure geometric fact has a close relation with
the effect, known in Physics as \emph{Berry Phase}. The integral curve $P(t),\,t\in[0,T]$ on the
Grassmann manifold defines two isomorphisms between the fibers $\pi^{-1}(P_0)$ and
$\pi^{-1}(P(T)\equiv P_1)$:
\newcommand{\berryg}{\ensuremath{\mathfrak{B}_1}}
\newcommand{\berry}{\ensuremath{\mathfrak{B}}}
\begin{enumerate}
\item
For a point $\sigma\in\pi^{-1}(P_0)$, let $\psi_\sigma(t),\,t\in[0,T]$ be the horizontal curve
such that $\psi_\sigma(0)=\sigma$ and $\pi(\psi_\sigma(t))=P(t),\,t\in[0,T]$. Define an
isomorphism $\mathfrak{B}_1:\pi^{-1}(P_0)\To\pi^{-1}(P_1)$ as $\berryg(\sigma)=\psi_\sigma(T)$
\item
For a point $\sigma\in\pi^{-1}(P_0)$, let $\varphi_\sigma(t),\,t\in[0,T]$ be the solution of the
equation \ref{lifted_schrodinger}, such that $\varphi_\sigma(0)=\sigma$. Define a mapping
$\berry:\pi^{-1}(P_0)\To\pi^{-1}(P_1)$ as $\berry(\sigma)=\varphi_\sigma(T)$.
\end{enumerate}
When the Hamiltonian curve $P(t)$ on the Grassmann manifold is closed: $P(0)=P(T)$, then the
isomorphism $\berry:\pi^{-1}(P_0)\To\pi^{-1}(P_0)$ is known as the \emph{Berry phase} and the
isomorphism $\berryg:\pi^{-1}(P_0)\To\pi^{-1}(P_0)$ is known as the \emph{geometric Berry phase}.

We call a Hamiltonian curve $P(t)\in\gr{m}{\hilb},\,t\in[0,T]$ \emph{geometric} if the corresponding
curve $\varphi(t)\in\isom,\,t\in[0,T]$, in the total space, which is the solution of the equation
\ref{lifted_schrodinger} is horizontal; that is for each $t\in[0,T]$ the tangent vector
$H(t)\circ\varphi(t)$ is horizontal. As the vertical component of this tangent vector is
$H(t)_{11}\circ\varphi$, the curve $P(t)$ is geometric if and only if $H(t)_{11}=0$ for each $t\in[0,T]$,
where $H(t)_{11}:X(t)\To X(t)$ is the upper left component in the decomposition
$$
H(t)=
\left[
\begin{array}{ccc}
H(t)_{11} & & -H(t)_{12}^*\\
          & & \\
H(t)_{12} & & H(t)_{22}
\end{array}
\right]:\hilb=X(t)\oplus X(t)^\perp\To\hilb=X(t)\oplus X(t)^\perp
$$
\begin{prop}
For any curve $Q:[0,T]\To\gr{m}{\hilb}$ there exists such a time-dependent Hamiltonian
$H^Q(t):\hilb\To\hilb,\,t\in[0,T]$ that the curve $Q$ is a Hamiltonian curve for $H^Q(t)$ and is
geometric.
\end{prop}
\begin{prf}
The tangent vector $\dot{Q}(t)$, for any $t\in[0,T]$ is an element of the tangent space of the
Grassmannian: $\dot{Q}(t)\in Hom(Q(t),Q(t)^\perp)$. Consider the time-dependent Hamiltonian
$$
H^Q(t)=
\left[
\begin{array}{ccc}
0 & & -\dot{Q}(t)^* \\
&&\\
\dot{Q}(t) & & 0
\end{array}
\right],\qquad t\in[0,T]
$$
it is clear that the curve $Q:[0,T]\To\gr{m}{\hilb}\cong\proj_m(\hilb)$ satisfies the Hamiltonian
(Schr\"{o}dinger) equation $\dot{Q}(t)=[H^Q(t),Q(t)]$. This implies that the curve $Q(t)$ is
Hamiltonian. At the same time, for any $t\in[0,T]$, the component $H^Q_{11}$ is automatically 0.
Therefore, the tangent vectors $H^Q\circ\varphi(t)$ for the points
$\varphi(t)\in\pi^{-1}(Q(t)),\quad t\in[0,T]$ are ``pure horizontal''.
\end{prf}
It is clear that for geometric curves on the base manifold, the two mappings $\berry$ and $\berryg$
coincide. In other words, for geometric curves the \textbf{geometric Berry phase} and the
\textbf{Berry phase} coincide.
\newpage
\bibliographystyle{amsplain}

\end{document}